\documentclass[twocolumn,showpacs,preprintnumbers,amsmath,amssymb]{revtex4}
\usepackage{graphicx}
\usepackage{dcolumn}
\usepackage{bm}
\begin{document}
\bibliographystyle{apsrev}


\title{Structural properties of screened Coulomb balls}

\author{M.~Bonitz$^1$}
\author{D.~Block$^2$}
\email{block@physik.uni-kiel.de}
\author{O.~Arp$^2$}
\author{V.~Golubnychiy$^1$}
\author{H.~Baumgartner$^1$}
\author{P.~Ludwig$^1$}
\author{A.~Piel$^2$}
\author{A.~Filinov$^1$}
\affiliation{$^1$ITAP, $^2$IEAP, Christian-Albrechts-Universit\"{a}t zu Kiel,
D-24098 Kiel, Germany}


\date{\today}

\begin{abstract}
Small three-dimensional strongly coupled charged particles in a spherical confinement potential arrange themselves in a nested 
shell structure. By means of experiments, computer simulations and theoretical analysis, it is shown that their structural properties 
depend on the type of interparticle forces. Using an isotropic Yukawa interaction, quantitative agreement for shell radii and occupation is obtained.
\end{abstract}
\maketitle
The recently discovered Coulomb balls \cite{arp04} are an interesting new object
for studying strongly coupled systems. Coulomb balls consist of hundreds of
micrometer sized plastic spheres embedded in a gas plasma. The plastic spheres attain a high electric charge $Q$ of the order of several thousand elementary charges and arrange themselves into a highly ordered set of nested spherical shells with hexagonal order inside
the shells. Coulomb balls are a special form of 3D-plasma crystals
\cite{Goree,Zuzic,hayashi}. The formation of ordered clusters with nested shells was
also observed in laser-cooled trapped ion systems, e.g. \cite{itano,Drewsen},
and is expected to occur in expanding neutral plasmas \cite{pohl04,killian04}.

The same kind of ordering was found in molecular dynamics (MD) simulations, e.g. \cite{hasse91, tsuruta93, ludwig-etal.05pre} and references therein. In particular, the transition to the macroscopic limit \cite{totsuji02,schiffer02}, the symmetry properties of the individual shells including a Voronoi analysis \cite{tsuruta93} and metastable
intrashell configurations \cite{ludwig-etal.05pre,arp-etal.jp05} have been analyzed. Very large systems of trapped ions show a transition to the crystal structure of bulk 
material, which was measured by laser scattering \cite{bollinger}. 

Although the shell structure of ion crystals is quite well understood in terms of simulation results, these systems do not yet allow for systematic experimental studies of the structure 
inside the shells and the detailed occupation numbers of individual shells. The advantage of studying Coulomb balls is the immediate access to the full three-dimensional structure of
the nested shell system by means of video microscopy. The tracing of each individual particle is ensured by the high optical transparency of the system, which results from particle diameters of typically 5 ${\rm \mu m}$ at interparticle spacings of 500 ${\rm \mu m}$. Compared to atomic particles, the very high mass of the microparticles used here slows down all dynamic phenomena to time scales ranging from 10~ms to seconds. Therefore,
studies of Coulomb balls complement investigations of ion crystals, where dynamical studies are difficult.

Coulomb balls are in a strongly coupled state, i.e. the Coulomb coupling
parameter,  $\Gamma = \frac{Q^2}{{a} k_BT}$, where $a$ is the mean interparticle distance, attains large values ($\Gamma \gg 100$). Contrary to ion crystals, where the particles interact via the pure Coulomb force, the microparticles in a Coulomb ball are expected to interact by a Yukawa type potential, $V_{dd}=Q\,e^{-r\lambda_D}/r$, which was 
verified experimentally in complex plasmas \cite{Konopka}. Therefore, Coulomb balls are characterized by two parameters, the coupling parameter $\Gamma$ and the Debye shielding length of the plasma  $\lambda_D$. It is the intention of this paper to study the influence of
shielding on the structure of Coulomb balls, in particular, to pin down the differences from systems with pure Coulomb interaction. This will be done by comparing computer simulations with experimental results. At the same time, a study of spherical crystals with Yukawa interaction opens up an interesting new field which in a natural way bridges the gap between the above mentioned theoretical investigations of finite size Coulomb systems and 
the theory of macroscopic Yukawa plasmas, e.g. \cite{dubin,fortov03}.

\begin{figure}[h]
\includegraphics[height=4.5cm,clip=true]{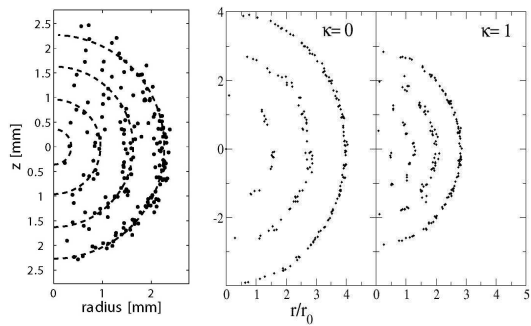}
\vspace{-0.5cm}
\caption{Radial particle distribution for $N=190$ given in cylindrical coordinates. Left: experiment \cite{arp04},right two figures: simulation results with Coulomb ($\kappa=0$), 
and Yukawa ($\kappa=1$) potential. The length unit in the right two figures is $r_{oc}$,
given by Eq. (\ref{eq:r0}).}
\label{fig:rd_190}
\end{figure}

{\em Experiment.} 
The experiment is described in detail in Refs. \cite{arp04,arp-etal.jp05,arp05}, so
here we only summarize the main results from a systematic investigation of 43 Coulomb balls 
consisting of 100 to 500 particles. All Coulomb balls were trapped under identical experimental conditions. All of them had a spherical shape and their diameter was in the range
of 4-5\,mm. A typical experimental result for a cluster and its shell structure is shown in the left part of Fig.~\ref{fig:rd_190}. In all 43 Coulomb balls  a similar shell structure was observed and  the shell radii $R_s$ and the shell occupation numbers $N_s$ were measured. 
Further, from the pair correlation function the typical mean interparticle distance was derived, which for all $N$ was close to $a\simeq 0.6$\,mm. The mean intershell distance $d$ was found to be $d=(0.86\pm0.06) a$, which yields a ratio $d/a\approx 0.82$ in good agreement with local icosahedral ordering \cite{hasse91}. An important experimental result is that the intershell distance is constant over the whole Coulomb ball and implies a constant average density of particles and ions, which is equivalent to a parabolic electric potential well used for the simulations below. A more detailed experimental verification of the parabolic confinement well will be described elsewhere \cite{arp05}. A different case with 'self-confinement' of a dust cloud in a strongly anharmonic potential was recently discussed in \cite{totsuji05}. 

{\em Simulations.}
For a theoretical explanation of the experimental results we have performed molecular dynamics (MD) and thermodynamic Monte Carlo (MC) simulations using the Hamiltonian 
\begin{equation}
H=\sum_{i=1}^N \left\{ \frac{p_i^2}{2 m} + U_c({\bf r}_i) \right\}
+\frac{1}{2}\sum_{i\ne j} V_{dd}({\bf r}_i - {\bf r}_j).
\label{eq:h}
\end{equation}
We assume that the Coulomb balls consist of particles with same mass and charge and that a stationary state is reached close to thermodynamic equilibrium. Furthermore, the observed isotropic particle configuration suggests to use an isotropic interaction potential. Screening effects are included in static approximation using Debye(Yukawa)-type 
pair potentials $V_{dd}$ given above. In the simulations we use dimensionless parameters, with lengths given in units of the ground state distance of two particles, $r_{0c}$, defined in Eq.~(\ref{eq:r0}), hence in this paper $\kappa =r_{0c}/\lambda_D$. In experimental papers, $\kappa=a/\lambda_D$ is often used. In accordance with the experiment on Coulomb balls \cite{arp05} and previous experiments and simulations on ion crystals \cite{dubin}, we use a screening-independent confinement potential $U_c(r)=m\omega^2\cdot r^2/2$. As a result, in our model the configuration of the Coulomb balls is determined by three parameters: 
particle number $N$, screening parameter $\kappa$ and temperature $T$. Since experimental plasma densities and temperatures are not precisely known, we have performed a series of calculations for different values of $\kappa$ and $T$. Furthermore, a wide range of particle numbers, up to $N=503$, has been analyzed.

{\em Results.} 
Consider first the theoretical ground state configurations ($T=0$) in the case of Coulomb interaction, $\kappa=0$, which were obtained by classical MD simulations using an optimized simulated annealing technique \cite{ludwig-etal.05pre}. Using about $1000$ independent runs for each value of $N$ ensured that the ground state is reached. In addition, we have performed MC simulations in the canonic ensemble with a standard Metropolis algorithm, which
allows for a rigorous account of finite temperature effects. Both simulations yield identical configurations at low temperature. Fig.~\ref{fig:rd_190} shows a comparison of MD simulation and experiment for the case of $N=190$ particles. In both cases four concentric spherical shells are observed, which are the result of a balance between confinement potential $U_c$ and interparticle repulsion $V_{dd}$. 

\vspace{-0.4cm}
\begin{figure}[h]
\includegraphics[height=7.5cm,clip=true,angle=-90]{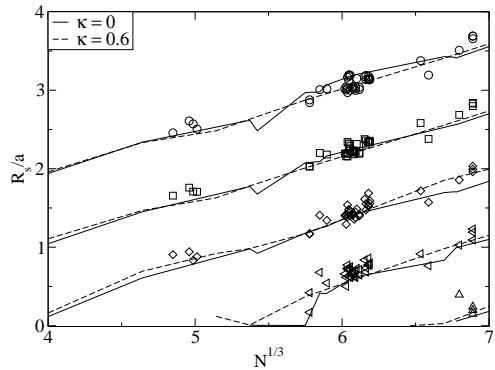}
\vspace{-0.15cm}
\caption{Experimental (symbols) and simulation (lines) results for the shell 
radii of three-dimensional Coulomb balls in units of the mean interparticle distance.
}
\label{fig:rs_n13}
\end{figure}

For a more detailed quantitative comparison between experiment and simulation we analyze 
the dependence of the shell radii $R_s$ on the cluster size $N$ (Fig.~\ref{fig:rs_n13}).
The interparticle distance $a$ serves as a common length scale as it is accessible in experiment and simulation. There is an overall increase $\propto N^{1/3}$ of the experimental $R_s$ for all shells and all 43 analyzed clusters. Exceptions occur around values of $N$ where new shells emerge. The same behaviour is obtained from the MD simulations. 
Without any free parameter a very good agreement of experimental radii and  Coulomb MD results (full lines) is observed, in particular concerning the absolute values, the slope and the equidistance of the shells. Further, this results holds even in case of 
a Yukawa potential if $\kappa$ is small (dashed lines in Fig.~\ref{fig:rs_n13}). These findings already imply two things: First, the approach to use an screening independent parabolic confinement potential to model the experiments is justified. This is a marked difference from a situation with self-confinement, which does not yield equidistant shells \cite{totsuji05}. Second, the general scaling of the shell radii of weakly shielded Coulomb balls $\propto N^{1/3}$ is the same as for pure Coulomb systems, such as ion crystals.

However, a marked difference between experiment and simulations of pure Coulomb systems is observed for the shell population numbers $N_1\dots N_4$. Table~\ref{tab:ns_190} shows the shell population numbers for various screening parameter $\kappa$ of a Coulomb ball with $N=190$ as obtained from MD simulations and experiment. Clearly, the MD results yield systematically more particles in the outer part of the cluster for $\kappa=0$ than observed in experiment. Further, Tab.~\ref{tab:ns_190} shows that, with increasing $\kappa$, particles move from the outer shell inward. Interestingly, for $\kappa=0.58\dots 0.63$, the simulations yield exactly the same shell configuration as the experiment. Hence, the different population numbers may be attributed to the influence of screening and hence weakening of the interaction potential.

\begin{figure}[h]
\includegraphics[height=10cm,clip=true,angle=-90]{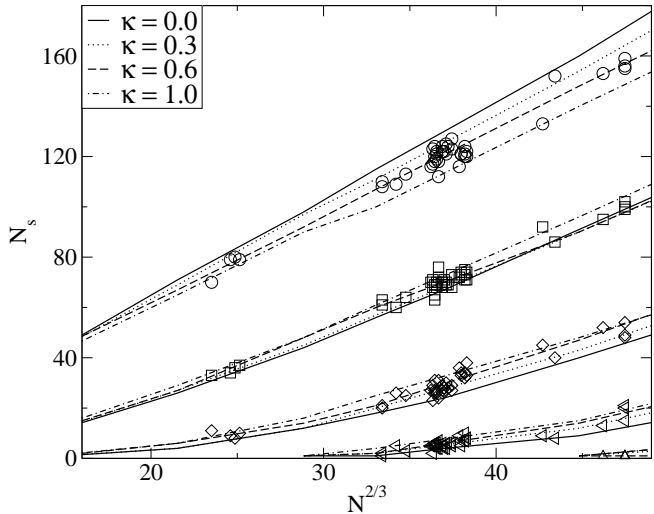}
\vspace{-0.3cm}
\caption{Experimental (symbols) and simulation (lines) results for the shell 
population of three-dimensional Coulomb clusters.
}
\label{fig:ns_n23}
\end{figure}

To investigate this in more detail, this comparison was extended to all 43 Coulomb balls. 
Due to their different size and even different number of shells the systematic differences in shell population of Coulomb and Yukawa systems can be studied comparing the experimental results and systematic MD-simulations. The result is shown in Fig.~\ref{fig:ns_n23}. For all systems Coulomb, Yukawa as well as the experimental data the shell population of all shells shows an almost linear behavior as a function of $N^{2/3}$. However, the experimentally 
obtained shell population of the outermost shell $N_4$ is always smaller than those of a Coulomb system (solid line) and  the inner shells show a systematically higher population. 
From the Yukawa MD-simulations (dashed lines) this tendency is confirmed. It is clearly found that particles move to inner shells with increasing $\kappa$. Hence, the finding discussed for the Coulomb balls with $N=190$ in Tab.~\ref{tab:ns_190} holds generally. This tendency reflects the fact that, from an energetic point of view, the higher population of inner shell
becomes less costly due the shielding than the occupation of the outermost shell, where the confinement by the trap dominates the potential energy.     

In more detail, we find that the outermost shell exhibits the largest absolute change with $\kappa$ and, therefore, it is best suited for a detailed comparison with the experimental data, see Fig.~\ref{fig:ns_n23}. From a best fit to the experimental data, we find a screening parameter $\kappa^{\rm EXP}=0.62 \pm 0.23$. An independent analysis for the other shells confirms this result, e.g. the third shells, yield $\kappa^{\rm EXP}=0.58 \pm 0.43$. 
Determining the mean interparticle distance $a$ from the first peak of the pair distribution function $\kappa^{\rm EXP}$ translates into an average Debye length $\lambda_D/a=1.54\pm 0.7$. Furthermore, as one can see in the right hand part of Fig.~\ref{fig:rd_190}, an increase of $\kappa$ leads to compression of the entire cluster, which is due to the reduction of the potential $V_{dd}$. The fact that more and more particles move from the outer shells inward has the consequence that closed shell configurations are already reached for a smaller number  $N^*$ of total particles compared to $N^*_c$ in the Coulomb case. 
While for $\kappa=0$, the first closed shell is found at 12 particles, for $\kappa\gtrsim 4.7$ the ground state of a cluster with 12 (and 11 as well) particles contains one particle 
in the center and $N^*=10$. For $\kappa=0.6$ closure of the 2nd to 4th shell is observed for $N^*_2=54, N^*_3=135 , N^*_4=271  $, compared to the Coulomb case where $N_{2c}^*=57, 60$ \cite{tsuruta93}, $N_{3c}^*=154$ \cite{arp-etal.jp05} and $N_{4c}^*=310$, cf. also Fig.~\ref{fig:rs_n13}.

\begin{table}
\begin{tabular}[t]{|c|c|c|c|c|c|c|c||c|}
\hline
$\kappa \rightarrow$   & $0$     & $0.2$   & $0.3$ & $0.4$ & $0.5$   & $0.6$ & $1.0$ & Experiment\\ [0.3ex]
\hline
$N_1$ & $1$    & $1$   & $2$   & $2$   & $2$    & $2$   & $4$   & $2$\\ [0.2ex]
$N_2$ & $18$   & $18$  & $20$  & $20$  & $21$   & $21$  & $24$  & $21$\\ [0.2ex]
$N_3$ & $56$   & $57$  & $57$  & $58$  & $58$   & $60$  & $60$  & $60$\\ [0.2ex]
$N_4$ & $115$  & $114$ & $111$ & $110$ & $109$  & $107$ & $102$ & $107$\\ [0.2ex]
\hline
\end{tabular}
\caption{Experimental (last column) and theoretical shell configuration of the Coulomb ball $N=190$. $N_1\dots N_4$ denote the particle numbers on the i-th shell beginning in the center. 
}
\label{tab:ns_190}
\end{table}

After analyzing the shell populations we now consider the shell width. The larger 
roughness of the shells in the experiments, cf. Fig.~\ref{fig:rd_190}, is attributed to small anisotropies of the experimental confinement and finite depth resolution of the imaging equipment as well as temperature effects. While the measurements are at room temperature, the simulations refer to $T=0$. Therefore we have analyzed the influence of temperature on the shell radii and populations with the help of MC simulations. From the results we conclude that the effect of temperature on the shell configurations $N_s$ is negligible for $\kappa=0.6$. At constant finite $T$ we find that an increase of $\kappa$ leads to a reduction of shell roughness. Contrary to that, a temperature increase at elsewhere fixed parameters in fact leads to a roughening of the shells proportional to $\sqrt{T}$ for the outer shell and an even stronger effect for the inner shells. This tendency will become evident from the analytical results below.

{\em Analytical results.} 
The main influence of screening on Coulomb balls is readily understood with the help of analytical results, which can be found for $N=2$. First, the ground state distance $r_0(\kappa)$ follows from minimizing the potential energy U in Eq.~(\ref{eq:h}):
\begin{eqnarray}
\frac{e^{\kappa r_0}r_0^3}{1+\kappa r_0} = \frac{Q^2}{\frac{m}{2}\omega^2} 
\equiv r^3_{0c}.
\label{eq:r0}
\end{eqnarray} 
Eq.~(\ref{eq:r0}) yields the two-particle distance, $r_{oc}$, in an unscreened system as a function of $r_0$ and is easily inverted numerically \cite{approx_r0}. The ratio $r_{0}/r_{0c}$ is always smaller than unity and monotonically decreasing with $\kappa$, 
thereby confirming the above observation of screening-induced compression of the Coulomb balls. Second, we analyze the cluster stability by expanding the potential $U$ in terms of small fluctuations, $y\equiv r-r_0$, around the ground state, up to second order: 
$U(r)-U(r_0)=\frac{1}{2}U''(r_0)y^2 \equiv \frac{m}{2}\Omega^2y^2$. This defines an 
effective local trap frequency $\Omega$
\begin{eqnarray} 
\label{eq:omega2}   
\Omega^2(\kappa)&=& 3\omega^2\left(1+\frac{1}{3}\frac{\kappa^2r_0^2}{1+\kappa r_0}\right) = \frac{3}{m}\frac{Q^2}{r^3_0}f_2(\kappa), \quad
\\
\nonumber
f_2(\kappa)&=& e^{-\kappa r_0}\left( 1+\kappa r_0+\kappa^2r_0^2/3 \right)\quad,
\end{eqnarray} 
which allows us to estimate the width of the Coulomb ball shells. Finally, we 
compute the variance of the particle distance fluctuations, $\sigma_r$, for particles 
in a parabolic potential with frequency $\Omega$ at temperature $T$ and obtain 
$\sigma^2_r=4 k_BT /(m\Omega^2)$. This allows for two interesting conclusions, 
which explain the above MC results: At constant screening, the shell width grows with temperature as $\sqrt{T}$ while screening reduces the shell width. One might be tempted to conclude that increased screening makes particle transitions between neighboring shells less likely and thus stabilizes the cluster against melting. However, the opposite is true, because screening also reduces the distance between shells which is of the order of $r_0$.
The relative importance of both tendencies can be discussed in terms of the {\em relative distance fluctuations}, a critical value of which determines the onset of radial melting (Lindemann criterion).
\begin{eqnarray} \label{eq:ur}
u_r^2 \equiv \frac{\sigma^2_r}{r^2_0} &=&
\frac{4}{3}\frac{1}{\Gamma_2^*}, \qquad 
\Gamma_2^*=\Gamma_2 f_2(\kappa)
\quad.
\end{eqnarray} 
$u_r$ is related to an effective coupling parameter, 
$\Gamma_2^*$ which depends on the interaction strength of two trapped particles -- via the Coulomb-type coupling parameter, 
$\Gamma_2\equiv Q^2/(k_BT r_0)$, and on the screening strength -- via the function $f_2(\kappa)$. 
$f_2$ monotonically decreases with $\kappa$ ($u_r$ increases), thus {\em screening always destabilizes the Coulomb balls}.  

Finally, these analytical results are closely related to those for macroscopic homogeneous Yukawa systems, e.g. \cite{dubin,fortov03}. This limit is recovered by replacing, in (\ref{eq:omega2}), $r_0$ by the mean interparticle distance $a$ at a given density $n$, $a=(3n/4\pi)^{1/3}$. Then the local trap frequency becomes $\Omega^2 \rightarrow \omega^2_{pd}f_2({\kappa})$, showing that, in a Coulomb system, $\Omega$ approaches the dust plasma frequency $\omega_{pd}$ whereas, in the case of  screening, the result is modified by a factor $\sqrt{f_2(\kappa)}$ \cite{one_half}. Also, the effective coupling parameter $\Gamma_2^*$ is in full analogy to the macroscopic result \cite{fortov03}. 

In summary, we have presented a combined experimental, numerical and theoretical 
analysis of small spherical charged particle clusters. The excellent experimental accessibility of these systems has been demonstrated. The structure of these clusters deviates from models with pure Coulomb interaction and requires the inclusion of 
static screening. For the particle number range $N=100 \dots 500$, comparison with 
the MD and MC simulations has allowed us to determine the screening parameter averaged over the clusters as $\lambda_D/a \approx 1.5$. These Coulomb balls are representative for finite Yukawa systems, combining shell properties observed in spherical Coulomb clusters with screening effects found in Yukawa plasmas. Since the shell occupation numbers have now been critically analyzed, our results confirm earlier conclusions about the shell structure of ion clusters, where such an analysis was not accessible yet. The results are relevant for other strongly correlated charged particle systems, such as crystal formation of droplets in expanding laser produced plasmas, where shielding becomes important.

\begin{acknowledgments}
This work is supported by the Deutsche Forschungsgemeinschaft via SFB-TR 24 grants 
A3, A5 and A7 and, in part, by DLR under contract 50WM0039. We acknowledge discussions 
with W.D. Kraeft and M. Kroll's assistance in conducting the experiments.
\end{acknowledgments} 


\begin{thebibliography}{26}
\bibitem{arp04}
O.~Arp, D.~Block, A.~Piel, and A.~Melzer, Phys. Rev. Lett. {\bf 93}, 165004 (2004)
\bibitem{Goree} J.B.~Pieper, J.~Goree, and R.A.~Quinn, Phys. Rev. E {\bf 54}, 5636 (1996).
\bibitem{Zuzic} M.~Zuzic et al.
Phys. Rev. Lett. {\bf 85}, 4064 (2000)
\bibitem{hayashi}  Y.~Hayashi, Phys. Rev. Lett. {\bf 83}, 4764 (1999)
\bibitem{itano} D.J. Wineland, J.C. Bergquist, W.M. Itano, J.J. Bollinger, and
C.H. Manney, Phys. Rev. Lett. {\bf 59}, 2935 (1987)
\bibitem{Drewsen} M.~Drewsen et al.,
Phys. Rev. Lett. {\bf 81}, 2878 (1998)
\bibitem{pohl04} T.~Pohl, T.~Pattard, and J.M.~Rost, Phys. Rev. Lett. {\bf 92},
155003 (2004)
\bibitem{killian04} T. Killian, Nature {\bf 439}, 815 (2004)
\bibitem{hasse91} R.W.~Hasse, and V.V.~Avilov, Phys. Rev. A {\bf 44}, 4506 (1991)
\bibitem{tsuruta93} K.~Tsuruta, and S.~Ichimaru, Phys. Rev. A {\bf 48}, 1339 (1993)
\bibitem{ludwig-etal.05pre} P.~Ludwig, S.~Kosse, and M.~Bonitz,
Phys. Rev. E {\bf 71}, 046403 (2005)
\bibitem{totsuji02} H.~Totsuji et al., Phys. Rev. Lett. {\bf 88}, 125002 (2002)
\bibitem{schiffer02} J.P.~Schiffer, Phys. Rev. Lett. {\bf 88}, 205003 (2002)
\bibitem{arp-etal.jp05} O.~Arp et al., J. Phys. Conf. Series {\bf 11}, 234 (2005)
\bibitem{bollinger} W.M.~Itano et al., Science {\bf 279}, 686 (1998)
\bibitem{Konopka} U.~Konopka, G.E.~Morfill, and L.~Ratke, Phys. Rev. Lett. {\bf 84}, 891 (2000)
\bibitem{dubin} D.H.E.~Dubin, and T.M. O'Neill, Rev. Mod. Phys. {\bf 71}, 87 (1999)
\bibitem{fortov03} V.E.~Fortov, et al., Phys. Rev. Lett. {\bf 90}, 245005 (2003)
\bibitem{arp05} O.~Arp, D.~Block, and A.~Piel, to be published
\bibitem{totsuji05} H.~Totsuji, C.~Totsuji, T.~Ogawa, and K.~Tsuruta, Phys. Rev. E {\bf 71}, 045401 (2005)
\bibitem{approx_r0} A useful analytical approximation 
for $r_{0}$ in Eq. (\ref{eq:r0}) as a function of $r_{0c}$ is 
$x=x_c+\frac{\ln{(1+x_c)}-x_c}{x_c^2+3x_c+3}x_c(1+x_c)$, where $x=\kappa r_0$, 
which has an error of less than  $1\%$, for $x_c=\kappa r_{0c}<1.5$. 
\bibitem{one_half} This result 
differs slightly from the exact macroscopic result \cite{fortov03} [by the coefficient 
$1/3$ instead of $1/2$ in the last term in $f_2$] which is a consequence of performing 
this replacement in the two-particle expression (\ref{eq:r0}).
\bibitem{kraeft_to_be} H.~Baumgartner, W.D.~Kraeft, and M.~Bonitz, to be published
\end{thebibliography}

\end{document}